\documentclass [prl,twocolumn] {revtex4}
\usepackage{graphicx}
\usepackage{epsfig}
\begin {document}
\title {Vortices and phase transitions of interacting bosons in a rotating lattice ring}
\author{Qi Zhou\\ \textit{Physics Department, Ohio-State University, Columbus, OH, 43210} }
\date {\small\today}
 \begin{abstract}
 We investigate the ground state properties of the repulsive interacting bosons in an one dimensional rotating lattice ring, and reveal that the superfluid density of the system and the mass current it can carry in the rotating coordinate are periodic functions of the velocity of the lattice. In the weakly interacting limit, the vortices are generated at the critical velocities, and in the strongly interacting limit, the phase twist of the bosons induced by the rotating lattice manifests the superfluid density and drive the system to undergo phase transitions between the superfluid and Mott insulator.
 \end{abstract}
\maketitle
The response of the superfluid to the external rotation is of great interest to the condensed matter physicists. Unlike the normal fluid, a superfluid can only follow the external rotation when the rotation frequency exceeds a critical value so that vortices can come into the system. The development of cold atom physics provides us with unique opportunity to investigate the critical velocity of superfluid through its highly controllability in experiments. For example, an optical lattice with desired lattice depth, geometry and rotation frequency can be applied to the cold atomic gas. Recent experiments at JILA have found the interesting pinning phenomenon and structural transition of the vortex lattice in an imposed rotating two dimensional optical lattice\cite{Cornell}. A fundamental question that arises naturally is, when and how do vortices come into the superfluid when a periodic external potential is imposed? Moreover, it is known that, if the lattice barrier is increased, the superfluid will become weaker and weaker, and finally undergoes a phase transition from the superfluid to the Mott insulator at integer filling. This raises another question about the fate of the weak superfluid, can it survive under external rotation? Both of the two questions are not unique to cold atom physics, but also to the condensed matter physics, which involves the crystal environment.

\begin{figure}[bp]
\begin{center}
\includegraphics[width=3in]{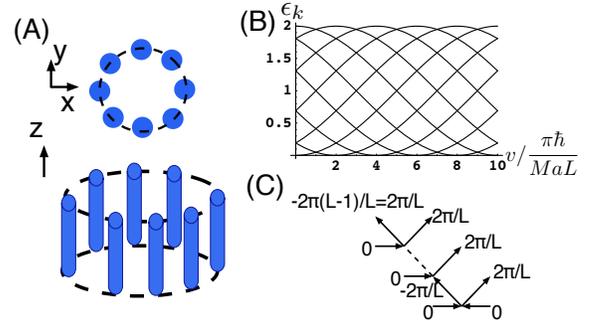}
\caption{Left: The schematic figure for the system. Right top: the energy spectrum for $L=10$. Right bottom: the interacting process scattering $L$ particles initially in $k=0$ state to $k=\frac{2\pi}{L}$ state.}\label{fig: system}
\end{center}
\end{figure}

In this letter, we consider the ground state of the repulsive interacting bosons in a rotating optical lattice ring. The ring geometry of the optical lattice has been proposed to be experimentally realizable by the interference of Laguerre-Gauss mode laser and a plane wave laser\cite{Luigi}. Their proposal can be easily extended to produce a rotating lattice ring. If we open a detuning $\Delta$ between the frequencies of these two modes, the lattice ring will rotate at the speed of $v\propto\frac{R\Delta}{L}$, where $R$ is the radius of the ring where the lattice sites locate and $L$ is the number of sites. As shown in Fig.(\ref{fig: system}A), we assume the confinement of the system in $z$ direction is quite weak, and in the dilute limit, the behaviour of the atoms are sitting in the ground state in $z$ direction . In the $x-y$ plane, the lattice sites are located along a ring, and the confinement in the radial direction is strong enough to quench the motion of the atoms along this direction. So we will concentrate on the physics along the azimuthal direction, and the system can be simplified as a one dimensional moving lattice with periodic boundary condition. 

We would like to point out that our work is different from recent work in \cite{Inguscio, Altman}, in which they studied the dynamical instability of a bosonic superfluid with high crystal momentum, which is not the ground state of the system. Here we will consider the evolution of the ground state as a function of the velocity of the lattice. The main results of our work are summarized as following: (I) Both the superfluid density $\rho_s$ and mass current $J$ in the rotating coordinate are periodic functions of the velocity $v$, i.e., $J(v)=J(v+T_v)$ and $\rho_s(v)=\rho_s(v+T_v)$, where $T_v=\frac{2\pi\hbar}{MaL}$, where $M$ is the mass of the bosons and $a$ is the lattice spacing. (II) In the weakly interacting limit, vortices are generated at the critical velocities $v_c=\frac{(2l+1)\pi\hbar}{MaL}$ through the Umklapp process, where $l$ is an integer number. (III) In the strong interacting limit, before the vortices come into the system, the rotating lattice drives the system from the superfluid to Mott insulator. If the rotation is continuously increased, the superlfluid phase comes back again with vortices\cite{Holland}. \\

\textbf{\textit{Hamiltonian}}: To construct the proper basis for the many body hamiltonian, we start from the single particle hamiltonian in the moving coordinate, $H=H_0-\vec{v}\cdot\vec{P}$, where $H_0=\frac{P^2}{2M}+V_{lat}$ is the hamiltonian in the stationary coordinate, $\vec{v}$ is the velocity of the moving frame, and $V_{lat}$ is the lattice potential, $V_{lat}(r)=V_{lat}(r+a)$. Hereafter, we use $r$ to represent the coordinate in the ring. As $H(r)=H(r+a)$, the solution of the Schr\"{o}dinger equation $H\psi(r)=E\psi(r)$ has the formula $\psi_{nk}(r)=e^{ikr}u_{nk}(r)$ where $u_{nk}(r)=u_{nk}(r+a)$ and $n$, $k$ is the band index and the crystal momentum respectively. To satisfy the periodic boundary condition in the ring, $\psi_{nk}(r+La)=\psi_{nk}(r)$, $k$ can only take the discrete values, $k=\frac{2\pi}{a}\frac{l}{L}, l=0,\pm{1},\pm{2}...\pm[L/2]$. Substitute it back to the Schr\"{o}dinger equation, we obtain
\begin{equation}
\left(\frac{(-i\nabla-M{v}+k)^2}{2M}+V-\frac{Mv^2}{2}\right)u_{nk}=\epsilon_{nk}u_{nk}.
\end{equation}
Compare it with the $v=0$ case, i.e., the equation for $u_nk$ in the stationary frame, we immediately know that $\epsilon_{nk}=\epsilon_{nk-Mv}^0-\frac{Mv^2}{2}$, where the superscript $0$ denotes the physics observable in the stationary coordinate. In tight binding limit, $\epsilon_{nk}^0=t_n(1-\cos{ka})$, we obtain
\begin{equation}
\epsilon_{nk}=t_n\left(1-\cos{(k-Mv)a}\right)-\frac{Mv^2}{2}.
\end{equation}
A typical energy spectrum $\epsilon_k$ within a single band as a function of $v$ is shown in Fig.(\ref{fig: system}).  

Now let us turn to the many body hamiltonian, 
\begin{equation}
H=\int{dr}\Psi^\dagger(\frac{P^2}{2M}-{v}\cdot{P}+V_{lat})\Psi+\int{dr}\Psi^\dagger\Psi^\dagger\Psi\Psi,
\end{equation}
where $g=\frac{4\pi\hbar^2a_{sc}}{M}\int{d}zd\omega{u_0^4}(z)h_0^4(\omega)$ is the effective one dimensional interaction. $a_{sc}$ is the scattering length,  $u_0(z)$ and $h_0(\omega)$ are the ground state wave functions along $z$ and radial direction. We can expand the single particle part of the Many body hamiltonian, $K=\int{dr}\Psi^\dagger(\frac{P^2}{2M}-v\cdot{P}+V_{lat})\Psi$, in the basis of $\psi_{nk}$, $\Psi=\sum_{nk}\psi_{nk}a_{nk}$, where $a_{nk}$ is the annihilate operator for the single particle eigenstate $\psi_{nk}$. In this basis, $K$ has a quite simple formula, $K=\sum_{nk}\epsilon_{nk}a^\dagger_{nk}a_{nk}$.  In the coordinate space, it becomes
\begin{equation}
K=-\sum_{n<i,j>}t_n(e^{i\frac{Mva}{\hbar}}b^\dagger_{ni}b_{nj}+c.c)+\sum_{ni}(t_n-\frac{Mv^2}{2})b^\dagger_{ni}b_{ni},
\end{equation}
where $b_{ni}=\frac{1}{\sqrt{L}}\sum_{k}e^{-ikR_i}a_{nk}$ is the annihilate operator for the Wannier function $W_{ni}(r)=\sum_ke^{-ikR_i}\psi_{nk}(r)$ which is localized at site $i$ and $<i,j>$ denotes the nearest neighbor sites. We can also expand the interaction energy $E_{int}=g\int{dr}\Psi^\dagger(r)\Psi^\dagger(r)\Psi(r)\Psi(r)$ in the  basis of Wannier function. If we only keep the onsite interaction within a single band in the tight binding limit, we obtain the effective many body hamiltonian,
\begin{equation}
H=-t\sum_{<ij>}(b^\dagger_ib_{j}e^{i\theta}+b^\dagger_{i}b_je^{-i\theta})+U\sum_i\frac{n_i(n_i-1)}{2}\label{Ha},
\end{equation}
where $\theta=\frac{Mva}{\hbar}$, $n_i=b^\dagger_ib_i$, $U=g\int{dr}|W_i(r)|^4/L^2$ . We have already dropped the constant term in the kinetic energy for simplicity.\\

\textbf{\textit{Small $U$ limit}}: From Fig.(\ref{fig: system}B), it is clear that, within each region $\frac{(2l-1)\pi}{L}<\frac{Mva}{\hbar}<\frac{(2l+1)\pi}{L}$, the single particle ground state  $\psi_{k_l}\sim{e}^{i\frac{2\pi{l}}{aL}r}$ carries a vortex with wilding number $l$, as the atom acquires a phase $2\pi{l}$ when it goes long the ring from $r=0$ to $r=La$ . Turn to a many body system with $N$ interacting particles, an interesting question arises immediately, how does the macroscopic wave function evolve from $\psi_{k_l}^N$ to $\psi_{k_{l+1}}^N$ when we increase the lattice velocity?  

To obtain the effective hamiltonian in the weakly interacting limit, we rewrite the hamiltonian in Eq.(\ref{Ha}) in the momentum space as,
\begin{equation}
H=\sum_k\epsilon_ka^\dagger_ka_k+\frac{U}{L}\sum_{k_1k_2k_3k_4}a^\dagger_{k_1}a^\dagger_{k_2}a_{k_3}a_{k_4}\delta_{k_1+k_2-k_3-k_4-\frac{2s\pi}{a}},
\end{equation}
where $s$ is an integer. Unlike the homogenous case, the difference between the total crystal momentum before and after the collision can be any reciprocal lattice vector $\frac{2s\pi}{a}$, which is well known as the Umklapp process in lattice.  For simplicity, we concentrate on the region $0<\frac{Mva}{\hbar}<\frac{2\pi}{L}$. In the weakly interacting limit, the ground state can be confined in the subspace composed by $|m\rangle=a^{\dagger{m}}_0a^{\dagger(N-m)}_1/\sqrt{m!(N-m)!}|0\rangle$, where $a_0^\dagger$($a_1^\dagger$) creates a particle in $k_{0}=0$($k_1=2\pi/L$) state. Because of the crystal momentum conservation, we notice $\langle{m'}|H|m\rangle\sim\delta_{m,m'\pm{L}}$. To scatter $L$ particle from $k_0$ state to $k_{1}$ state, we need incorporate the high order scattering process, and the one of the lowest order is shown in Fig.(\ref{fig: system}C). The effective hamiltonian is
\begin{equation}
H=\epsilon_0a^\dagger_0a_0+\epsilon_1a^\dagger_1a_1+g_0a^\dagger_0a_0a^\dagger_1a_1+g_1(a^{\dagger{L}}_1a^L_0+c.c),\label{em}
\end{equation}
where $g_0=\frac{U}{L}$ and $g_1=(\frac{U}{L})^{L-1}/{\prod_{i=2}^{i=L-1}(\frac{\epsilon_0+\epsilon_1}{2}-\epsilon_{i})}$. One might assume the last term in Eq.(\ref{em}) is a high order process and should be quite weak, but as we have seen, the kinetic energy is degenerate at $v_c$. This coupling term thus plays a non-perturbational role. 

Rewrite the wave function as $|G\rangle=\sum_mc_m|N/2+mL\rangle$, and project the Schr\"{o}dinger equation to the Fock state, we obtain an effective Schr\"{o}dinger equation for $c_m$, 
\begin{equation}
Ec_m=U(m)c_m-g_1(\frac{N}{2})^L\left(c_{m+1}+c_{m-1}\right),\label{sc}
\end{equation}
where $U(m)=\left((\epsilon_0-\epsilon_1)mL+(g_1'-g_0)m^2L^2\right)$ and $g_1'=g_1(\frac{N}{2})^{L-2}\frac{L}{2}$.  At sufficiently large filling,  i.e., $(\frac{Un}{t})^LL>\frac{Un}{t}$, it is straightforward to check that $g_1'$ will be greater than $g_0$. Consequently, the effective hamiltonian is an one dimensional harmonic oscillator and the ground state at $v_c$ where $\epsilon_0=\epsilon_1$ is a fragmental state
\begin{equation}
|G\rangle=\left(\frac{a^{\dagger{L}}_0-(-1)^La^{\dagger{L}}_1}{\sqrt{2L!}}\right)^{\frac{N}{L}}|0\rangle,
\end{equation}
whose single particle density matrix has more than one macroscopic eigenvalues, 
\begin{equation}
\left(
\begin{array}{cc}
\langle{a}^\dagger_0a_0\rangle&\langle{a}^\dagger_0a_1\rangle\\
\langle{a}^\dagger_1a_0\rangle&\langle{a}^\dagger_1a_1\rangle
\end{array}\right)=\frac{N}{2}\left(
\begin{array}{cc}
1&0\\
0&1
\end{array}\right).
\end{equation}
\begin{figure}[tbp]
\begin{center}
\includegraphics[width=2.2in]{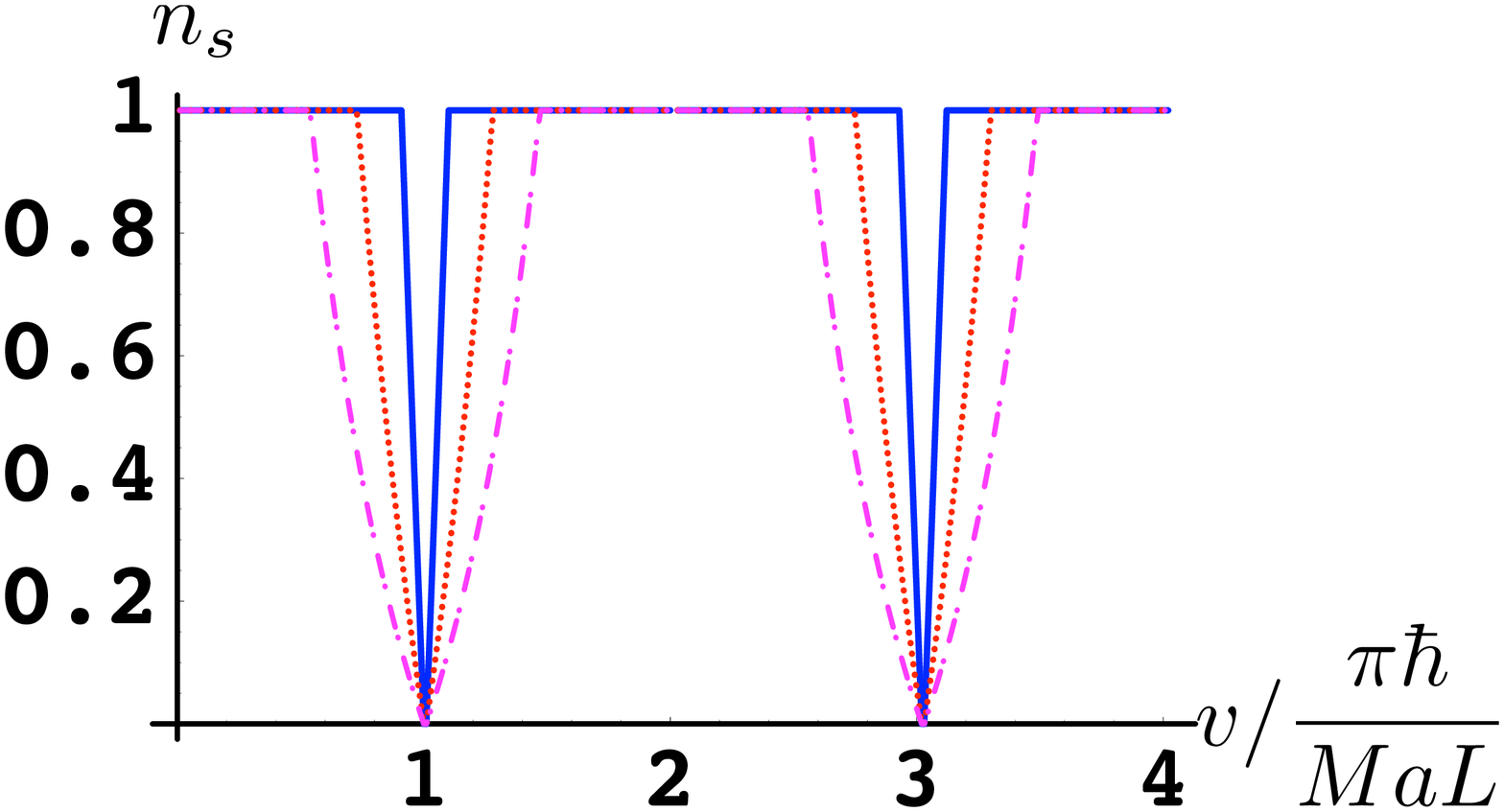}
\includegraphics[width=2.2in]{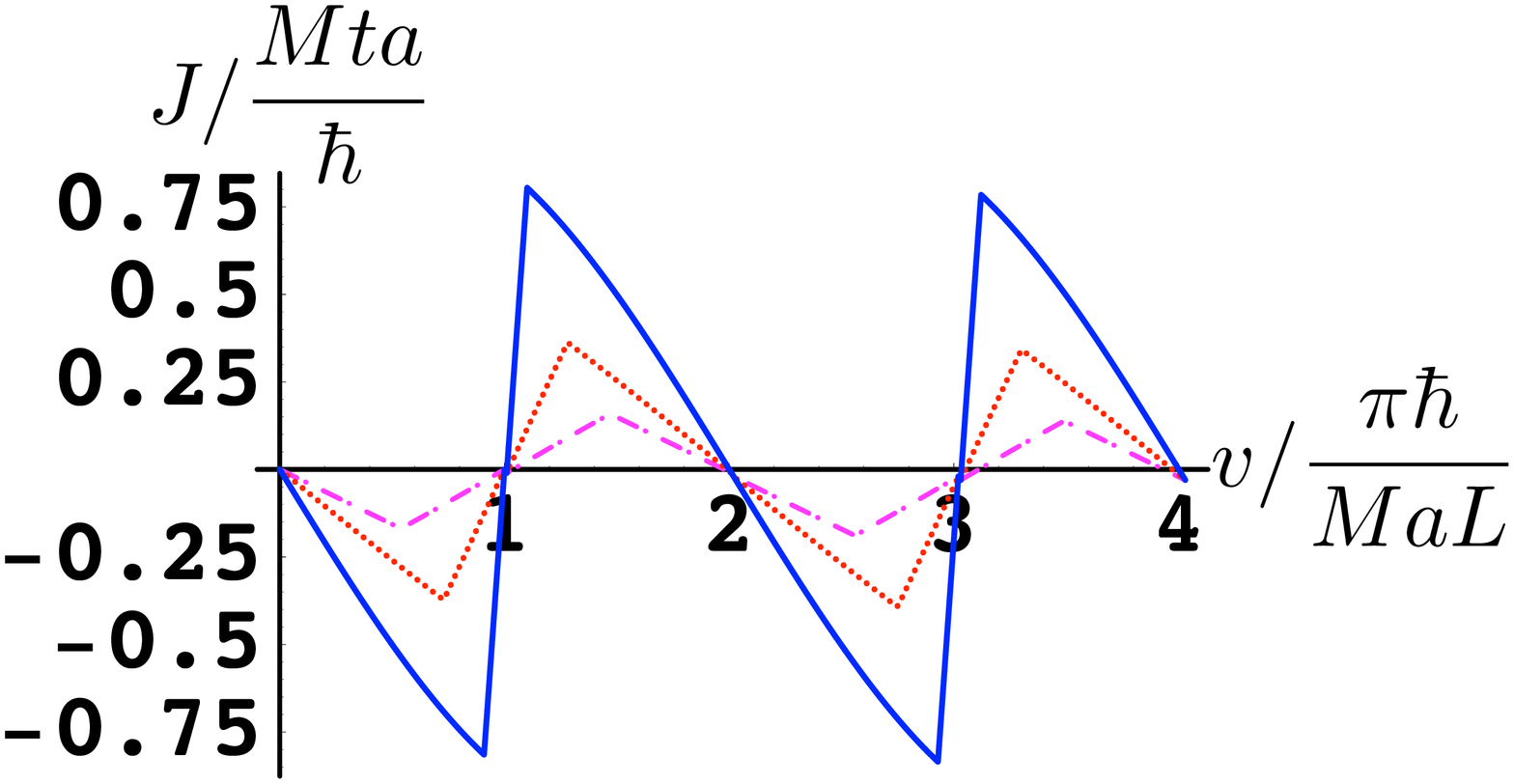}
\caption{Top: The superfluid fraction. Bottom: The mass current. The solid, dotted and dash-dotted line represent $L=3, 6$ and $10$, where $\frac{Un}{t}=0.9$, and $n=10^{5}$.}\label{fig:fig2}
\end{center}
\end{figure}When the velocity of the lattice is away from the critical value $v_c$,  Eq.(\ref{sc}) will be solved numerically to reveal how this fragmental state changes to the superfluid phases, which is reflected by evolution of the mass current and superfluid density. 

In the rotating frame, each atom in the eigenstate $\psi_{k_l}$ carries the mass current $J_{k_l}=\frac{M}{\hbar}\nabla_{k_l}\epsilon_{k_l}$ in the rotating frame, and a straightforward calculation shows the average mass current per particle is,
\begin{equation}
J=\frac{Mta}{\hbar}\left(\frac{\langle{a}^\dagger_0a_0\rangle}{N}\sin(-\theta)+\frac{\langle{a}^\dagger_1a_1\rangle}{N}\sin(\frac{2\pi}{L}-\theta)\right).
\end{equation}
From the mass current, we can also define the superfluid fraction $n_s$. As the normal part of bosons follow the lattice with a velocity $v$, the contribution to the non-zero mass current in the rotating frame totally comes from the superfluid fraction. When a vortex with wilding number $l$ is present, the phase twist of the superfluid between two nearest neighbor sites is $\phi=\frac{2\pi{l}}{L}$, and the mass current 
can be written as  $J={n_sMta}\sin(\phi-\theta)/\hbar$, and consequently, 
\begin{equation}
n_s=\frac{\hbar{J}}{Mta\sin(\phi-\theta)}\label{esd}.
\end{equation}
The numerical results are shown in Fig.(\ref{fig:fig2}). We see, at small lattice velocity, all the atoms stay at the state $\psi_{k_0}$ and the superfluid fraction keeps as 1. The mass current in the rotating frame is negative,  as the superfluid keeps stationary in the laboratory frame.  But when the lattice velocity approaches the first critical velocity $\frac{\pi\hbar}{MaL}$,  more and more atoms occupy the state $\psi_{k_1}$, and both the amplitude of the mass current and superfluid fraction decreases. When the lattice velocity crosses $v_c$, most atoms will stay in the state $\psi_{k_1}$, and the many body wave function gradually changes to $\psi_{k_1}^N$. The supefluid fraction comes back to 1 and a vortex enters the supefluid. If the velocity increases continuously, similar procedure repeats, vortices with different wilding number come into the system one by one.\\

\textbf{\textit{Large $U$ limit}}: We will show that in this limit, the external rotation can drive the system from the superfluid phase to the Mott insulator. At large $U$ limit, the number fluctuation in each lattice site is suppressed, and we can take the ansatz for the ground state,
\begin{equation}
|G\rangle=\prod_i(\alpha|n-1\rangle+\beta|n\rangle+\gamma|n+1\rangle),
\end{equation}
where $|\alpha|^2+|\beta|^2+|\gamma|^2=1$, and minimize the energy $\langle{G}|H-\mu{N}|G\rangle$ for the grand canonical ensemble. The chemical potential $\mu$ is fixed by the total particle number. Because of the presence of the external rotation, the order parameter $\varphi_m=\langle{b_m}\rangle$ aquires a phase, and we can write it as $\varphi_m=\varphi{e}^{im\phi}$, where $\varphi=|\langle{b_m}\rangle|$. We have assumed a uniform phase twist in the ring. To satisfy the periodic boundary condition, $\phi$ must satisfy $(L-1)\phi=-\phi$, i.e., $\phi=\frac{2\pi{l}}{L},l=0,\pm1,\pm2...\pm[\frac{L}{2}]$. For simplicity, we will call the system is in the $l$th branch if the phase $\phi$ is $\frac{2\pi{l}}{L}$. We will see immediately that the phase $\phi$ is not only the response of the superfluid to the external rotation, but also affects the amplitude of the superfluid density in turn. 

Apply the standard mean field procedure\cite{Fisher}, decoupling the tunneling part of the hamiltonian as,
\begin{equation}
b^\dagger_mb_{m+1}=\varphi_m^*{b}_{m+1}+\varphi_{m+1}{b}^\dagger_{m}-\varphi^*_m\varphi_{m+1},
\end{equation}
 we obtain the effective hamiltonian $H=K+\frac{U}{2}\sum_mn_m(n_m-1)$, where 
\begin{equation}
K=\sum_m-2t\cos(\theta-\phi)\varphi(e^{i\phi_m}b_m^\dagger+e^{-i\phi_m}b_m)+2t\cos(\theta-\phi)\varphi^2.\label{eH}
\end{equation}
\begin{figure}[tbp]
\begin{center}
\includegraphics[width=2.2in]{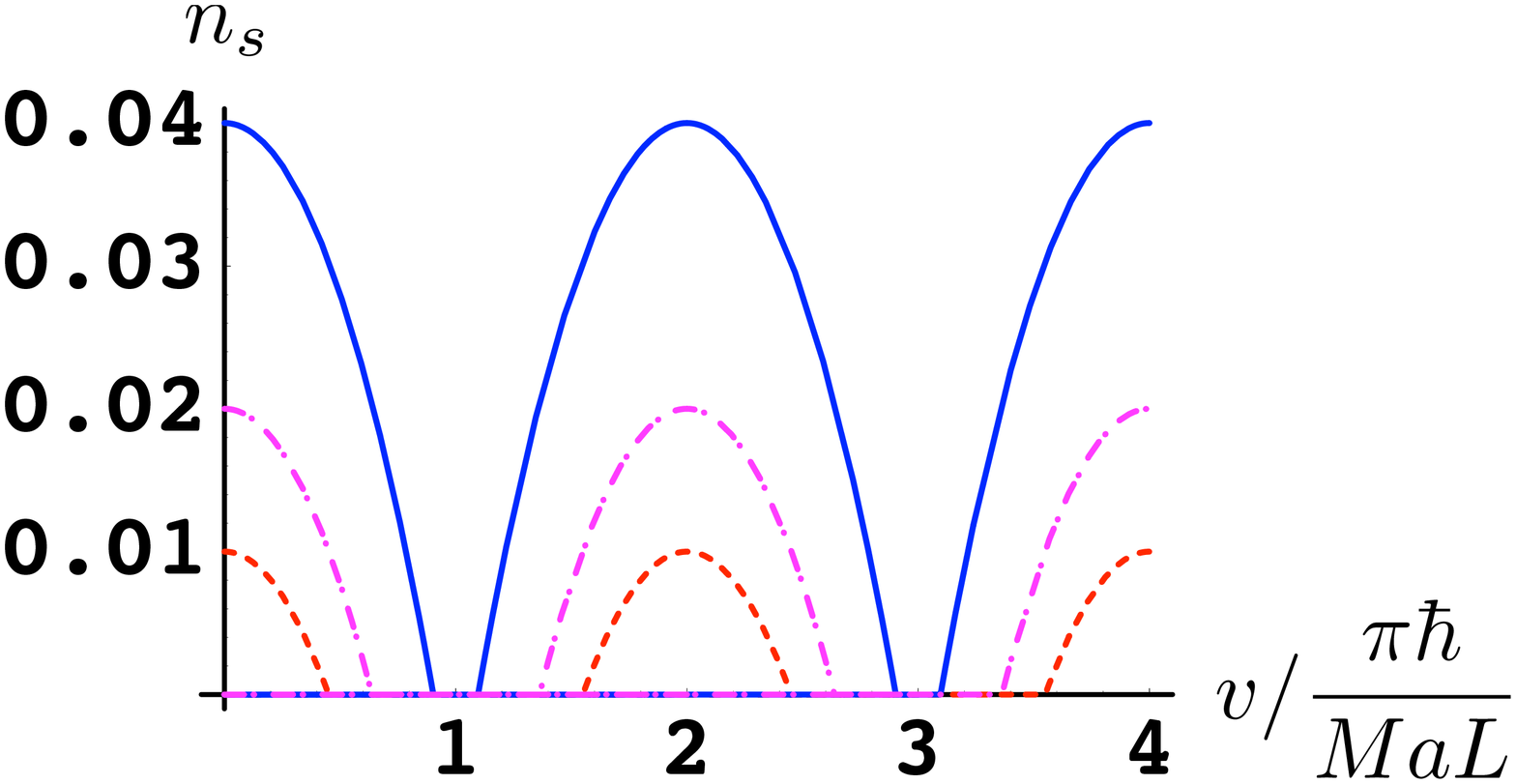}
\includegraphics[width=2.5in]
{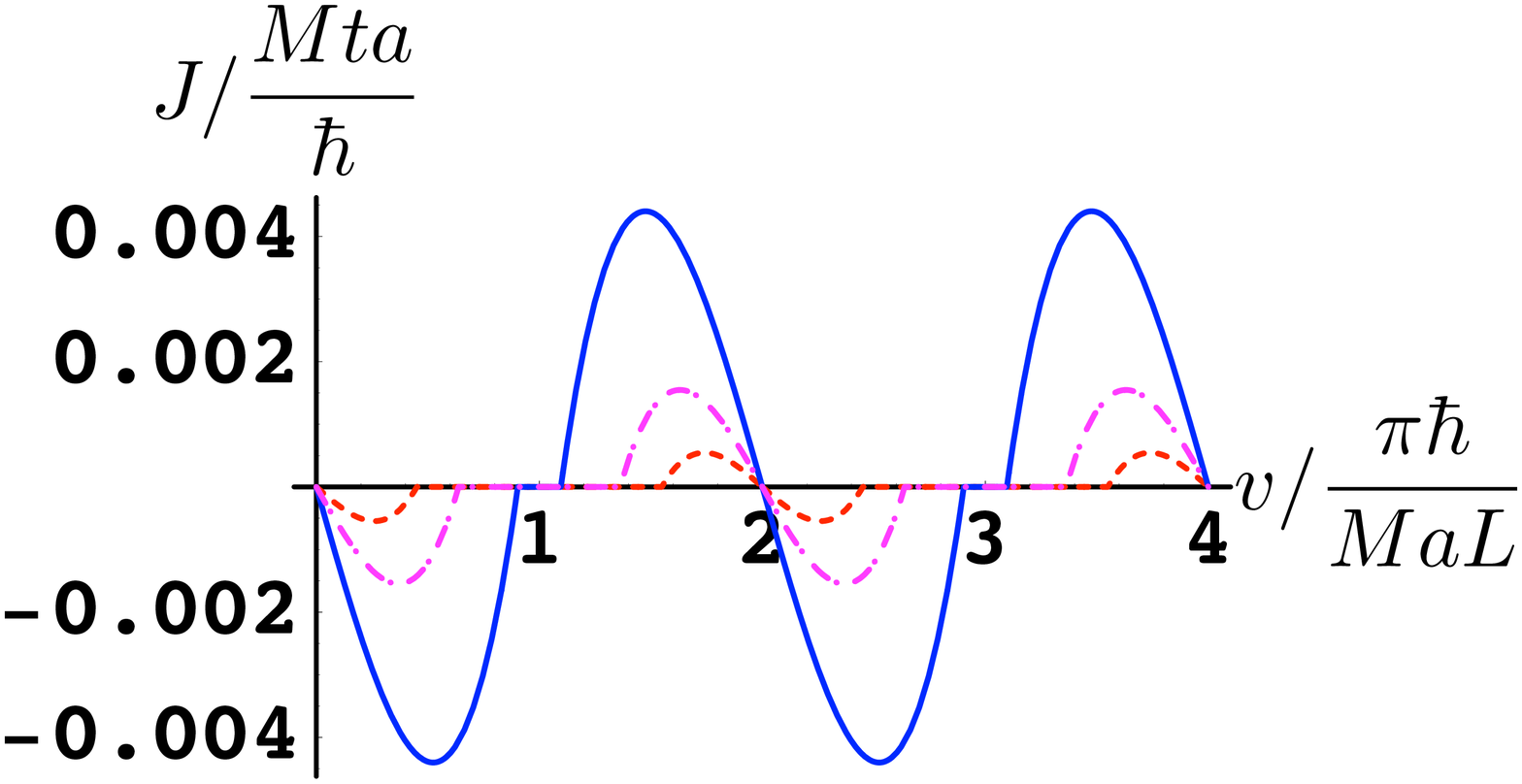}
\caption{Top: The superfluid fraction. Bottom: The mass current. The solid,  dash-dotted and dashed line represent $\frac{U}{U_c^0}=0.96, 0.98$ and $0.99$, where the number of sites is $L=10$, and the particle number per site is $n=1000$.}\label{fig:fig3}
\end{center}
\end{figure}Compared with the stationary case, it is clear that the tunneling is replaced by an effective one $t_{\mathrm{eff}}=t\cos(\theta-\phi)$. The factor $\cos(\theta-\phi)$ comes from the interference of  two different ways which can generate an atom at site $m$, say, from site $m-1$ and $m+1$, and it also implies the periodicity of the all the properties of the system. On the other hand, it is well known that, in the stationary lattice, with the increase of $U/t$, the superfluid density will decrease. Especially, at integer filling, a phase transition from superfluid to Mott insulator can occur. Consequently, \textit{the phase twist induced by the rotating lattice reduces the effective tunneling and manifests the superfluid density globally.} This is quite different from the well known phenomenon,\textit{ in a system without lattice the superfluid density can only be suppressed locally by rotation at the vortex core where the singularity of the phase occurs}. 

If we start from a stationary lattice, and adiabatically increase the velocity of the lattice, the superfluid initially staying in the $\phi=0$ state will vanish when $\theta$ satisfy $U=U_{c}^0\cos\theta$, where $U_{c}^0=t(2n+1+\sqrt{(2n+1)^2-1})$ is the critical interaction for the phase transition in the stationary coordinate. In another word, the transition happens when the velocity of the lattice reaches, 
\begin{equation}
v_M=\frac{\hbar}{Ma}\arccos\frac{U}{U_{c}^0}.
\end{equation}
From above equation, it is clear that $v_M$ can be much smaller than $v_c=\frac{\pi\hbar}{MaL}$ when $U$ is sufficiently large, i.e., $U>U_c^0\cos\frac{\pi}{L}$. In this large $U$ limit, before the vortex emerge, the system has already entered the Mott insulator phase. If we continuously increase the velocity of the lattice, not surprisingly, the superfluid fluid phase will come back again, as the effective hamiltonian is a periodic function of $v$. When $\frac{2\pi\hbar{l}}{MaL}-v_M<v<\frac{2\pi\hbar{l}}{MaL}+v_M$, the energy of $l$th branch becomes the lowest one, and the superfluid state with a vortex of wilding number $l$ is the ground state of the system. To quantitatively describe the system, we define the superfluid density as the usual way, $\rho_s=\varphi^2$, and consequently, the superfluid fraction $n_s=\rho_s/n$. According to Eq.(\ref{esd}), the mass current can also be calculated. The results are shown in Fig.(\ref{fig:fig3}) by solving the hamiltonian in Eq.(\ref{eH}). With the increase of the lattice velocity, $l=1$, $l=2$... branches come into the system one after another. Unlike the small $U$ limit, the Mott insulator plays the role as an intermediate state between different superfluid phases. \\

\textbf{\textit{Conclusion and remarks}}: We have investigated the ground state properties of the repulsive interacting bosons in a rotating lattice ring. An effective hamiltonian in the rotating frame is deduced and various phenomena about the vortices generation and phase transitions of the bosons are found. As we mentioned at the beginning of this paper, the behaviour of the superfluid in a crystal environment is generally a fundamental problem not only for the cold atoms community, but also for the condensed matter physics. As an example, the recent experiments on supersolid in a cylinder\cite{Moses1,Moses2} have attracted a lot of attention. Although the microscopic physics there is quite different, as the crystal order there is built upon the interactions among the Helium atoms themselves, not from the imposed external potential, 
some interesting phenomena found there are surprisingly similar as what we found here, i.e., the unit circulation of lattice for the generation of the vortex in the system and the destruction of the superfluid density by the external rotation. The relationship between these two systems, as well as the underlying physics, remains to be investigated.

We acknowledge helpful discussion with T.L Ho. This work is supported by NSF Grant DMR-0426149 and PHY-0555576.


\begin{thebibliography}{99}
\bibitem{Cornell}
S. Tung, V. Schweikhard, and E.A. Cornell,
Cond-mat/0607697 (2006)
\bibitem{Luigi}
Luigi Amico, Andreas Osterloh, and Francesco Cataliotti,
Phys. Rev. Lett 95, 063201(2005)
\bibitem{Inguscio}
L. Fallani, L. De Sarlo, J.E. Lye, M.Modugno, R.Saers, C.Fort, and M.Inguscio,
Phys. Rev. Lett 93, 140406(2004)
\bibitem{Altman}
E. Altman, A. Polkovnikov, E. Demler, B. I. Halperin, and M. D. Lukin,
Phys. Rev. Lett. 95, 020402 (2005)
\bibitem{Holland}
We noticed that a recent work, by R. Bhat, M. J. Holland, and L. D. Carr (Phys. Rev. Lett. 96, 060405), on a small system with a few bosons in a two dimensional rotating optical lattice made a similar claim on the superfluid to Mott insulator transition induced by the rotating lattice. But their approach is only valid in the small rotating frequency limit and can not be applied to the high rotating frequency to reveal the periodicity of all the properties of the system. 
\bibitem{Fisher}
M. P. A. Fisher, P. B. Weichman, G. Grinstein, and D. S. Fisher,
Phys. Rev. B 40, 546-570 (1989)
\bibitem{Moses1}
E.Kim and M.H.W. Chan,\ \
Nature 427, 255 (2004)
\bibitem{Moses2}
E.Kim and M.H.W. Chan,\ \
Science 305, 1941 (2004)

\end{thebibliography}
\end{document}